\documentstyle[11pt,newpasp,twoside,psfig]{article}
\newcommand{\ltapprox}{\raisebox{-0.5ex}{$\,\stackrel{<}{\scriptstyle\sim}\,$}}
\newcommand{\gtapprox}{\raisebox{-0.5ex}{$\,\stackrel{>}{\scriptstyle\sim}\,$}}

\begin{document}

\title{Properties of high-$z$ galaxies as seen through lensing clusters}
\author{Roser Pell\'o, Jean-Paul Kneib and Jean-Fran\c cois Le Borgne}\footnote{
And the {\it GRAVITATIONAL TELESCOPE } and
{\it LENSNET} Collaborations: TMR {\it Lensnet}
(http://www.ast.cam.ac.uk/IoA/lensnet) }
\affil{Observatoire Midi-Pyr\'en\'ees, LAT, UMR5572, 
Toulouse (France)}

\author{Bernard Fort and Yannick Mellier}
\affil{Institut d'Astrophysique de Paris, 
Paris (France)}

\author{Micol Bolzonella}
\affil{Istituto di Fisica Cosmica "G.P.S. Occhialini", 
Milano (Italy)}

\author{Luis Campusano}
\affil{Observatorio Cerro Cal\'an, Dept. de Astronom\' \i a, 
U. de Chile, 
Santiago (Chile)}

\author{Mireille Dantel-Fort}
\affil{Observatoire de Paris, DEMIRM, 
Paris (France)}

\author{Joan-Marc Miralles}
\affil{Astronomical Institute, Tohoku University, 
Sendai 
(Japan)}


\begin{abstract}
We discuss the first results obtained on the study of a sample of
high-$z$ galaxies ($2 \la z \la 7$), using the gravitational amplification
effect in the core of lensing clusters. Sources are located close
to the critical lines in clusters with well constrained mass
distributions, and selected through photometric redshifts,
computed on a large wavelength domain, and lens inversion techniques.
\end{abstract}


\section{Introduction}

The basis of our large collaboration program, involving different european institutions,
is to use clusters of galaxies as gravitational lenses to
build up and to study an independent sample of high-z galaxies. This sample
is important because it complements the large samples obtained in field surveys.
The idea is to take benefit from the large amplification factor close to the critical
lines in lensing clusters (typically 1 to 3 magnitudes) to study the
properties of the distant background population of lensed galaxies.
The signal/noise ratio in spectra of amplified sources
and the detection fluxes are improved beyond the limits of conventional techniques,
whatever the wavelength used for this exercise.
In particular, the amplification properties have been succesfully used
in the ultra-deep MIR survey of A2390 (Altieri et al. 1999), and the SCUBA
cluster lens-survey (Smail et al 98; Blain et al. 99).
This collaboration program is presently going on, and the next step is to
perform the spectroscopic follow up (mainly with the VLT) on a sample of high-z candidates 
selected through both photometric redshifts and lensing inversion procedures.

Number of high-z lensed galaxies have been found since the first case of lensed 
galaxy at $z \gtapprox 2$, the spectacular blue arc in Cl2244-02 (Mellier et al. 1991), 
and these findings strongly encourage our approach.
Among the recent examples of highly-magnified distant galaxies, identified either 
purposely or serendipitously, one can mention: the star-forming source
$\#384$ in A2218 at z=2.51 (Ebbels et al. 1996); the luminous z=2.7 arc
behind the EMSS cluster MS1512+36 (Yee et al. 1996; Seitz et al. 1998); the three z
$\sim$ 4 galaxies in Cl0939+47 (Trager et al. 1997); a z=4.92 system in
Cl1358+62 (Franx et al. 1997, Soifer et al. 1998); and the two red galaxies at
$z \sim 4$ in A2390 (Frye \& Broadhurst 1998, Pell\'o et al. 1999).

\section{The photometric redshift approach}

\begin{figure}
\psfig{figure=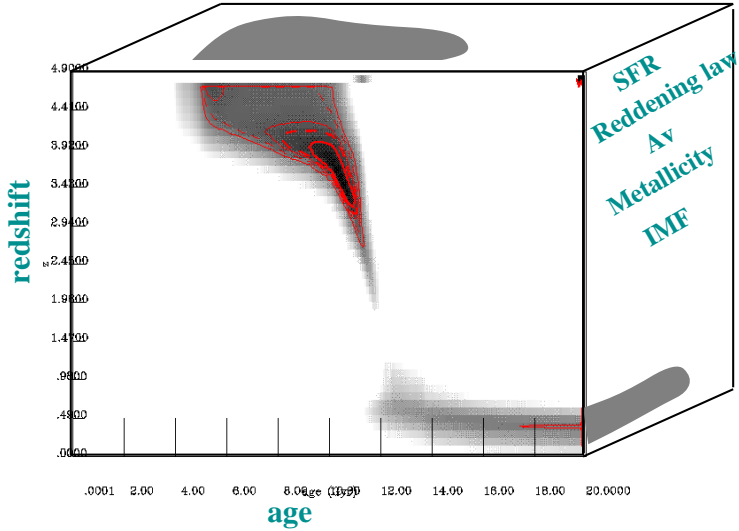,angle=270,height=7.0cm}
\caption{An artist view of the SED fitting procedure to compute {\it $z_{phot}$}}
\end{figure}

   Photometric redshifts (hereafter {\it $z_{phot}$} ) are computed using a standard SED 
fitting procedure originally developed by Miralles (1998). 
A new public version of this tool, 
called {\it hyperz}, is presently under developement (Bolzonella, Miralles and Pell\'o,
in preparation; see also Bolzonella \& Pell\'o, this conference). The set of templates 
includes mainly spectra from the Bruzual \& 
Charlot evolutionary code (GISSEL98, Bruzual \& Charlot 1993), and also a set of 
empirical SEDs compiled by Coleman, Wu and Weedman (1980) to represent the local 
population of galaxies. The synthetic database derived from Bruzual \& Charlot 
includes 255 spectra, distributed into
5 different star-formation regimes (all of them with solar metallicity): a burst of 0.1
Gyr, a constant star-formation rate, and 3 $\mu$ models (exponential-decaying SFR) with
characteristic times of star-formation chosen to match the present-day sequence of E,
Sa and Sc galaxies. The reddening law is taken from Calzetti (1997), with $A_V$ ranging
between 0 and 0.5 magnitudes. Flux decrements in the Lyman forest are computed
according to 
Madau (1995).
When applying {\it hyperz} to the spectroscopic samples available on the HDF,
the uncertainties are typically $\delta z / (1 + z) \sim 0.1$ (Bolzonella \& Pell\'o, 
this conference).

\section{Results and Future Developments}

\begin{figure}
\hbox{
\psfig{figure=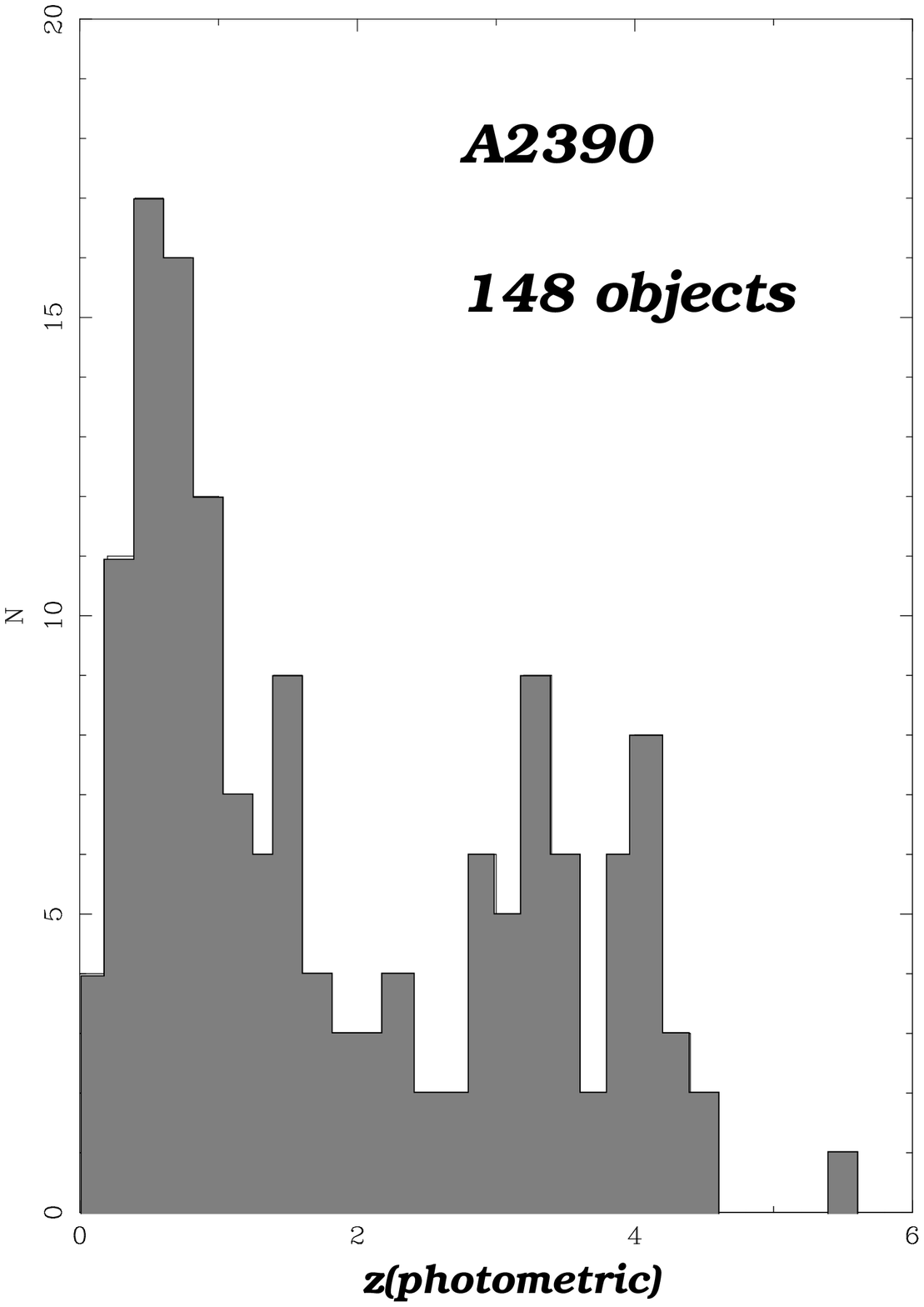,height=7.0cm}
\psfig{figure=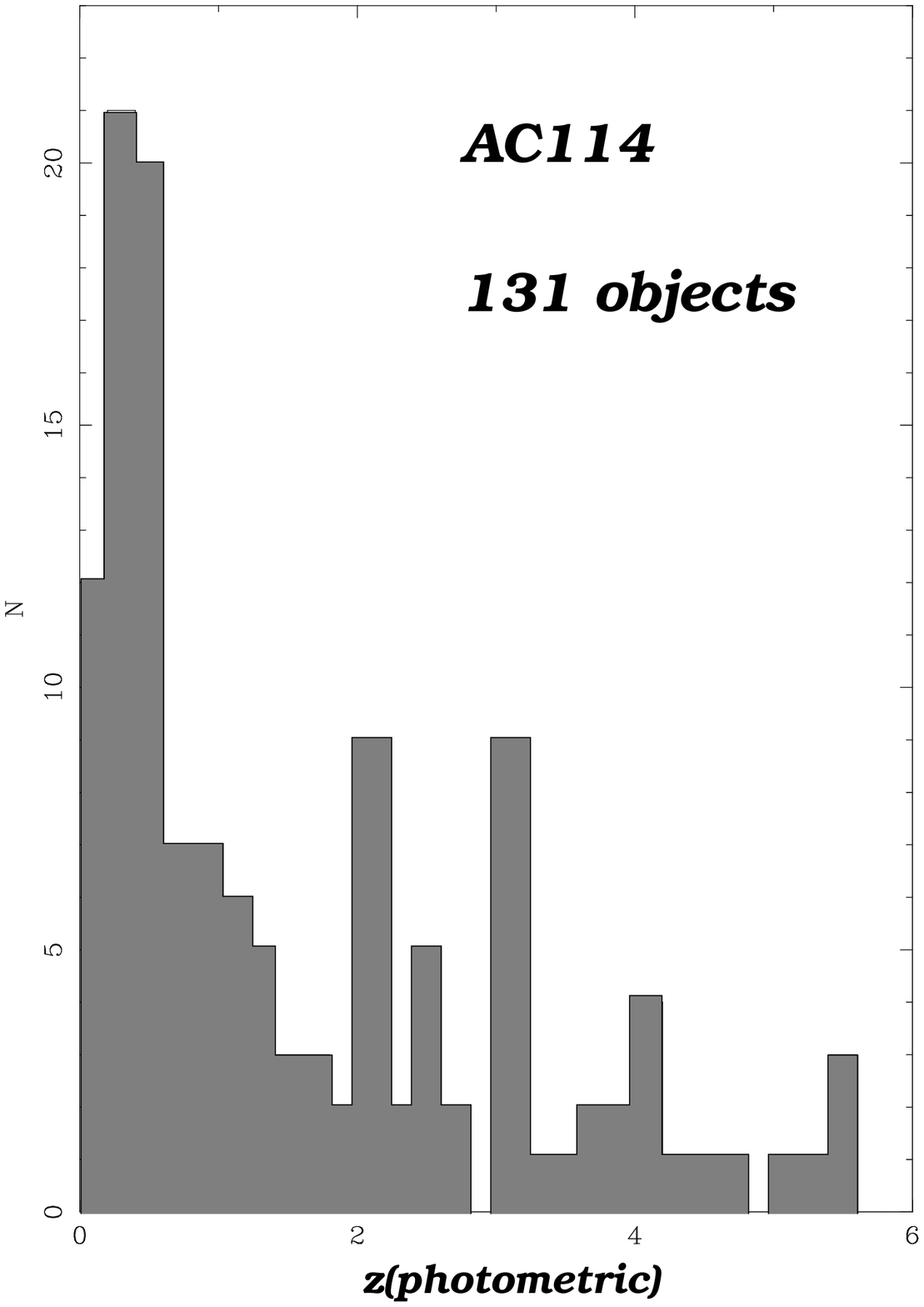,height=7.0cm}
\psfig{figure=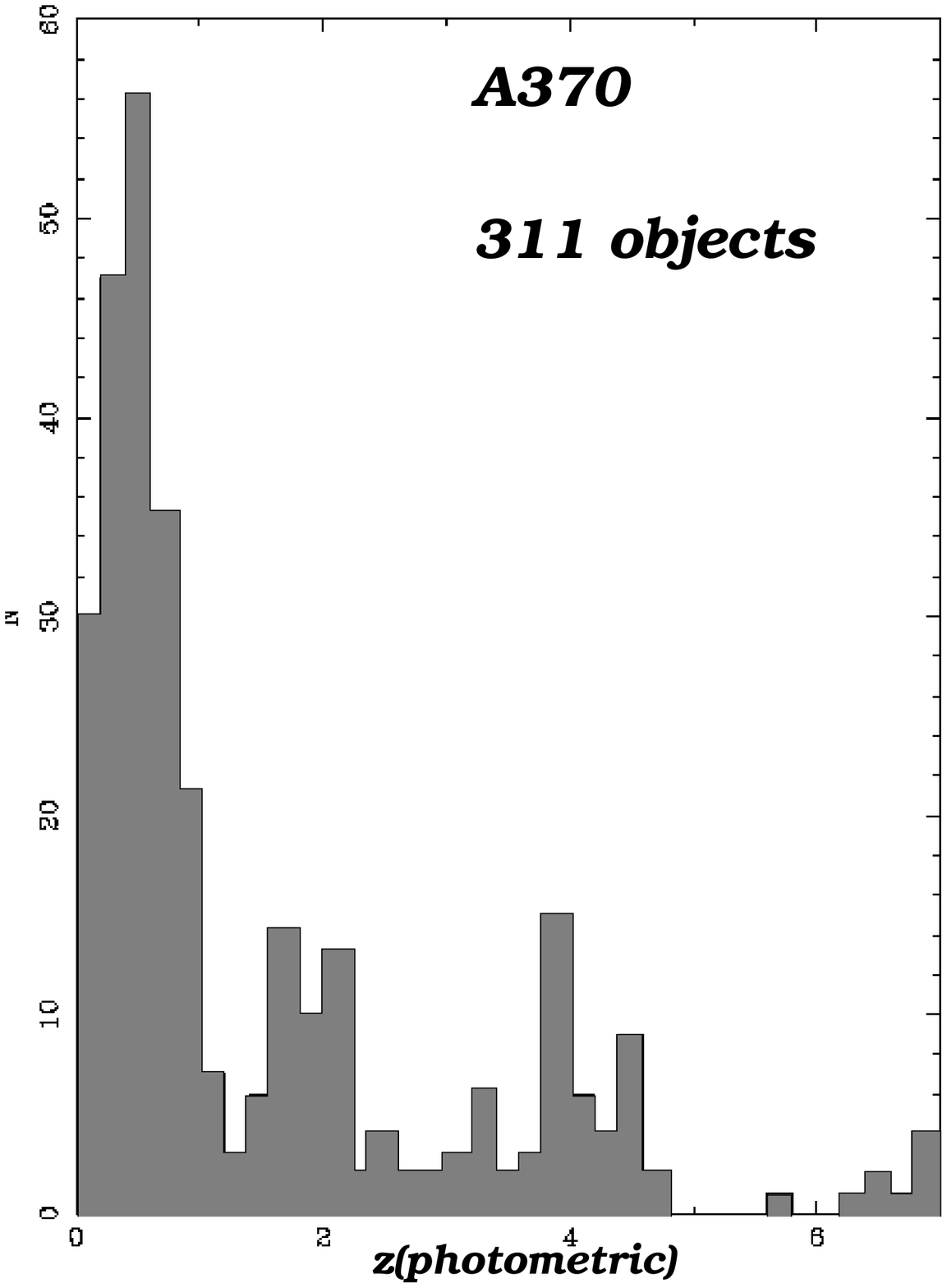,height=7.5cm}
}
\label{fig2}
\caption{{\it $z_{phot}$} distribution of arclets in A2390 (right), AC114 and
A370 (left), obtained with {\it hyperz}. In A2390, selection criteria
are based on the morphology of the WFPC2 images (elongation, orientation
and surface brightness); 
in A370, only a photometric selection was applied,
aimed to avoid the obvious bright cluster members;
in AC114, all objects within a 1.5 arcmin aperture from the center are displayed.}
\end{figure}

High-z lensed sources with {\it $z_{phot}$} $\ge 2$ are selected close to the appropriate 
critical lines. In all cases, {\it $z_{phot}$} are computed from broad-band photometry 
on a large wavelength interval, from B (U when possible) to K. 
This allows to cancel the biases which focus the convergence of the fitting procedure 
towards or against a particular type of galaxy or redshift domain, and also to reduce the
errors on {\it $z_{phot}$}.  Furthermore, it permits to optimize the instrument choice for
further spectroscopic follow-up (visible vs. near-IR bands).
The first spectroscopic surveys were performed on $\sim 4$m telescopes:
CFHT/ WHT/ ESO (3.6m, NTT) (Toulouse/Cambridge/Paris/Barcelona collaboration).
They have demonstrated the efficiency of the technique (see for instance Ebbels et al. 
1996, 1998, and Pell\'o et al. 1999). The present large VLT Program is focused on
an X-ray selected sample of 12 lensing clusters. Photometry was performed on
$\sim 4$m telescopes (NTT, 3.6m telescope ESO, ...), including
HST and/or other archive images when available. We intend to obtain the
whole spectroscopic follow up at VLT (FORS, ISAAC, ...).  The main goals of this 
program are: to determine the $z$ distribution
of a very faint subsample of high-$z$ lensed galaxies, unvisible otherwise; to study the SED of
$z \ge 1$ galaxies (especially  $z \ge 2$) for a sample less biased in
luminosity than the field (SFR history, permitted region in the age- metallicity-
reddening-... space) and to explore the 
dynamics of $z \ge 2$ sources by using 2D spectroscopy of arcs, a prospective issue
for future studies.

    Most of the photometric survey is presently completed. We have obtained the
(photometric) $z$ distribution of arclets in several well known clusters
(A2390, A370, Cl2244-02, AC114,...). Figure~2 displays
the {\it $z_{phot}$} distribution of arclets in three fields,
where the samples were defined according to different 
criteria. The typical number of high-z sources found in the inner 1' radius region of the 
cluster is $\sim 30$ to 50 at $1 \le z \le 7 $.
For a subsample of spectroscopically confirmed sources in different clusters (with
$0.4 \le z \le 4$), the
{\it $z_{phot}$} accuracy  has been checked as a function of the relevant parameters (SFR,
reddening, age and metallicity of the stellar population). We have also cross-checked
the consistency between the photometric, the spectroscopic and the lensing redshift
obtained from inversion methods (Ebbels et al. 1998). According to the present results,
the agreement between the three methods is good up to at least $z \ltapprox 1.5$. The 
comparison between the spectroscopic and the lensing redshift was already studied in 
the field of A2218 (Ebbels et al. 1998), and all the present results seem to follow 
this trend up to $z \ltapprox 1.5$ at least. Taking into account that {\it $z_{phot}$} and 
lensing inversion techniques produce independent probability distributions for amplified 
sources, combining both methods provides with a robust way to determine the redshift 
distribution of distant sources. This comparison gives promising results at $z \gtapprox 1.5$,
for the most amplified sources, 
but an enlarged spectroscopic sample is urgently needed to conclude, in particular
for the most distant candidates which could be the most distant galaxies
ever detected.

\begin{figure}
\psfig{figure=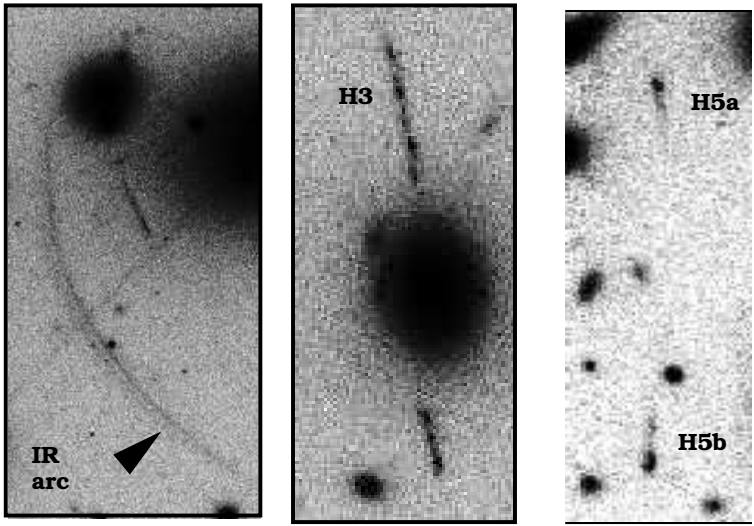,angle=270,height=7.0cm}
\caption{Several multiple images at $z \ge 1 $ in A2390, 
displayed in the I band (WFPC2).}
\end{figure}

The method is restricted to lensing clusters whose mass distribution is highly constrained 
by multiple images (revealed by HST or ground-based multicolor images), where the amplification
uncertainties are typically $\Delta m_{lensing}<$ 0.3 magnitudes.
Such well constrained mass distributions enable to
recover precisely the properties of lensed galaxies (morphology,
magnification factor). It is worth to note that highly magnified sources are 
presently the only way to access the dynamical
properties of galaxies at $z \ge 2$, through 2D spectroscopy, at a spatial resolution
$\sim$ 1 kpc. The two multiple-images at the same $z \sim 4$, observed behind
A2390, are un example of these reconstruction capabilities (Pell\'o et al. 1999).
Thanks to the lensing inversion, lensing clusters can
therefore be used  to calibrate photometric redshifts as well, up to the faintest 
limits in magnitude for a given z. They could be also used advantageously 
to search for primeval galaxies, in order to put strong constraints on the 
scenarios of galaxy formation.

\acknowledgments
We are grateful to G. Bruzual, S. Charlot, R.S. Ellis, S. Seitz 
and I. Smail for useful discussions on this particular technique.
Part of this work was supported by the TMR {\it Lensnet} ERBFMRXCT97-0172
(http://www.ast.cam.ac.uk/IoA/lensnet).

\end{document}